# Effect of average number of reference speckles on the SNR of object retrieved using off-axis holographic technique

**Abhijit Roy and Maruthi M. Brundavanam**

Department of Physics, Indian Institute of Technology Kharagpur, Kharagpur, India, 721302

E-mail: abhijitphy302@gmail.com, bmmanoj@phy.iitkgp.ernet.in

**Abstract**
The propagation of coherent light through a random scattering medium generates speckle pattern. Although the speckle is random in nature, the object information is scrambled into it. The scrambled object information can be retrieved using off-axis holographic technique where the retrieval is made from coherence hologram which is constructed from the interferogram formed by object speckle and a tilted reference speckle. It is shown that the SNR of the object retrieved from the coherence hologram, calculated from the far-field intensity correlation of the interferogram, depends on the average number of reference speckles. The effect of average number of reference speckles on the SNR of the retrieved object in off-axis holographic technique is investigated experimentally by changing the reference speckle while keeping the object speckle unchanged. The experimental observations are presented in this paper.

## 1. Introduction

The propagation of a coherent beam of light through a random scattering medium destroys the spatial coherency of the beam by distorting its wavefront. The unaffected temporal coherency allows the distorted wavefronts to interfere which results in random distribution of intensity, known as the speckle [1]. Although the spatial distribution of intensity of speckle is random, it is observed that due to its memory effect, the information of the input coherent beam is not completely lost rather scrambled into it [2]. Different techniques such as intensity correlation [1], wavefront shaping [3], and holography [4] is proposed and experimentally demonstrated to retrieve the object information from the speckle.

Recently, the focusing effect of scattered light inside a complex medium and its application in imaging tiny particles is demonstrated by controlling the wavefront of input coherent beam [5]. In another work, a technique based on the angular correlation method along with an iterative approach is demonstrated for non-invasive imaging through a random scattering medium [6]. In a recent development, the angular correlation technique is improved with the introduction of intensity correlation of speckle for imaging of objects through a scattering medium and is extended for imaging through a multi-fibre system [7, 8]. Recently, the observation of memory effect for double passage of light through a random scattering medium has paved the way to recover the information of an object buried deep inside a scattering medium and has also enabled to track its movement [9, 10]. In another separate work, imaging through a random scattering medium using bi-spectral analysis along with intensity correlation technique is demonstrated [11]. In most of the reported studies second order intensity correlation along with different phase retrieval algorithms is used for retrieving object information from speckle. However, it is also shown that higher order intensity correlation technique can also be employed for phase retrieval and apply it for imaging through random phase screen [12, 13]. Speckle imaging technique is also widely used in different types of biomedical studies and ghost imaging [14-17].

The techniques used in the discussed works for recovery of the lost phase of the complex coherence function for imaging of object through a random phase screen, is either phase retrieval algorithm or higher order intensity correlation technique, which are expensive either computationally or experimentally. Recently, a technique based on off-axis holography in association with the second order intensity correlation approach is proposed and demonstrated to recover the phase lost in the recorded speckle pattern and to reconstruct the complex coherence function [18]. This technique is employed to recover various types of 2D and 3D objects through different kinds of scattering media [19-21]. The technique is also useful in measuring the complex generalized stokes parameters [22], and determining the vorticity of vortex beams [23] through a random phase plate. In another work, it has been shown that speckle holographic technique can also be used for non-destructive testing such as detection of cracks, debonding, and delamination etc. of the sub-surface [24].

In the off-axis holographic technique proposed in Ref. [18], an object speckle is superposed with a tilted reference speckle to form a hologram which is used for different kinds of applications employing second order intensity correlation approach. Although the technique is used for different purposes, few effort is made to understand the technique fully [25]. In this paper, it is shown that the characteristic of reference speckle plays a major role in the object information retrieval process using off-axis holographic technique. The effect of average number of reference speckles on the second order intensity correlation function and on the SNR of the retrieved object is investigated by exploiting the far-field intensity correlation of the superposed speckle, and the experimental observations are presented.

## 2. Theoretical Background

Let us consider that the electric field vector, $\mathbf{E_O}(\mathbf{r},t)$ of a monochromatic object random field can be written in terms of the constituent electric field components $E_{Ox}(\mathbf{r},t)$ and $E_{Oy}(\mathbf{r},t)$ as

$$\mathbf{E_O}(\mathbf{r},t) = E_{Ox}(\mathbf{r},t)\,\hat{\mathbf{x}} + E_{Oy}(\mathbf{r},t)\,\hat{\mathbf{y}} \tag{1}$$

where $\mathbf{r}$ is position vector in the transverse observation plane, t is time, and $\hat{\mathbf{x}}$, $\hat{\mathbf{y}}$ are the two mutually orthogonal unit vectors. The random field can be characterized using coherence-polarization matrix or CP matrix. The CP matrix, $\Gamma(\mathbf{r_1},\mathbf{r_2})$ of a random field, $\mathbf{E}(\mathbf{r},t)$ is written as

$$\Gamma(\mathbf{r_1},\mathbf{r_2}) = \begin{bmatrix} \langle E_x^*(\mathbf{r_1})\,E_x(\mathbf{r_2})\rangle & \langle E_x^*(\mathbf{r_1})\,E_y(\mathbf{r_2})\rangle \\ \langle E_y^*(\mathbf{r_1})\,E_x(\mathbf{r_2})\rangle & \langle E_y^*(\mathbf{r_1})\,E_y(\mathbf{r_2})\rangle \end{bmatrix} \tag{2}$$

where '< >' denotes the ensemble average, and $\mathbf{r_1}$, $\mathbf{r_2}$ are the two spatial position vectors on the transverse plane. The spatial degree of coherence, $\gamma(\mathbf{r_1},\mathbf{r_2})$ of the random field can be calculated using the CP matrix discussed in Eq. (2) and also from the far-field intensity correlation following Ref. [26, 27] as

$$\gamma^2(\mathbf{r_1},\mathbf{r_2}) = \frac{\mathrm{tr}[\Gamma(\mathbf{r_1},\mathbf{r_2})\,\Gamma^\dagger(\mathbf{r_1},\mathbf{r_2})]}{|\mathrm{tr}[\Gamma(0)]|^2} = \frac{\langle \Delta I(\mathbf{r_1})\Delta I(\mathbf{r_2})\rangle}{\langle I(\mathbf{r_1})\rangle\langle I(\mathbf{r_2})\rangle} \tag{3}$$

where 'tr' denotes the trace of the matrix and $\Delta I(\mathbf{r}) = I(\mathbf{r}) - \langle I(\mathbf{r})\rangle$ is the spatial deviation of intensity from its mean value.

The CP matrix of object random field, $\mathbf{E_O}(\mathbf{r},t)$ is written as

$$\Gamma^O(\mathbf{r_1},\mathbf{r_2}) = \begin{bmatrix} \Gamma_{xx}^O(\mathbf{r_1},\mathbf{r_2}) & \Gamma_{xy}^O(\mathbf{r_1},\mathbf{r_2}) \\ \Gamma_{yx}^O(\mathbf{r_1},\mathbf{r_2}) & \Gamma_{yy}^O(\mathbf{r_1},\mathbf{r_2}) \end{bmatrix} \tag{4}$$

where the matrix elements are given by, $\Gamma_{ij}^O(\mathbf{r_1},\mathbf{r_2}) = \langle E_i^{O*}(\mathbf{r_1})\,E_j^O(\mathbf{r_2})\rangle$.

The object random field is made horizontally polarized for experimental simplicity. In this case, all other elements in $\Gamma^O(\mathbf{r_1}, \mathbf{r_2})$, except $\Gamma^O_{xx}$ is zero and the intensity correlation is found to be

$$\langle \Delta I(\mathbf{r_1})\, \Delta I(\mathbf{r_2}) \rangle = |\Gamma^O_{xx}(\mathbf{r_1}, \mathbf{r_2})|^2 \tag{5}$$

It can be observed from Eq. (5) that from the intensity correlation, the modulus of the complex coherence function can be retrieved. The retrieval of phase is essential to construct the complex coherence function, which is used to recover the object information scrambled in the random field. In order to retrieve the phase information, in the off-axis holographic approach, the object random field, $\mathbf{E_O}(\mathbf{r},t)$ is superposed with a reference random field, $\mathbf{E_R}(\mathbf{r},t)$, and the resultant field is given by

$$\mathbf{E}(\mathbf{r},t) = \mathbf{E_O}(\mathbf{r},t) + \mathbf{E_R}(\mathbf{r},t) \tag{6}$$

The CP matrix of reference random field, $\mathbf{E_R}(\mathbf{r},t)$ is written as

$$\Gamma^R(\mathbf{r_1}, \mathbf{r_2}) = \begin{bmatrix} \Gamma^R_{xx}(\mathbf{r_1}, \mathbf{r_2}) & \Gamma^R_{xy}(\mathbf{r_1}, \mathbf{r_2}) \\ \Gamma^R_{yx}(\mathbf{r_1}, \mathbf{r_2}) & \Gamma^R_{yy}(\mathbf{r_1}, \mathbf{r_2}) \end{bmatrix} \tag{7}$$

As the two random fields are experimentally generated from two independent scattering media, the mutual correlation between the two random fields can be taken as zero. And under the assumption of spatial stationarity and ergodicity of the fields, the spatial degree of coherence for the superposed random field can be derived as

$$\gamma^2(\mathbf{r_1}, \mathbf{r_2}) = \frac{\langle \Delta I(\mathbf{r_1}) \Delta I(\mathbf{r_2}) \rangle}{\langle I(\mathbf{r_1}) \rangle \langle I(\mathbf{r_2}) \rangle} = \frac{\mathrm{tr}[\Gamma(\mathbf{r_1}, \mathbf{r_2})\, \Gamma^\dagger(\mathbf{r_1}, \mathbf{r_2})]}{|\mathrm{tr}[\Gamma(0)]|^2} \tag{8}$$

where the $\Gamma(\mathbf{r_1}, \mathbf{r_2})$ is given by, $\Gamma(\mathbf{r_1}, \mathbf{r_2}) = \Gamma^O(\mathbf{r_1}, \mathbf{r_2}) + \Gamma^R(\mathbf{r_1}, \mathbf{r_2})$ which is also reported in our earlier work in Ref. [20].

The elements of the CP matrix, $\Gamma^R(\mathbf{r_1}, \mathbf{r_2})$ for a diagonally polarized reference random field is assumed to be $\Gamma^R_{xx} = \Gamma^R_{xy} = \Gamma^R_{yx} = \Gamma^R_{yy} = \Gamma_R$ and the resultant matrix $\Gamma(\mathbf{r_1}, \mathbf{r_2})$ is modified as

$$\Gamma(\mathbf{r_1}, \mathbf{r_2}) = \begin{bmatrix} \Gamma^O_{xx}(\mathbf{r_1}, \mathbf{r_2}) + \Gamma_R(\mathbf{r_1}, \mathbf{r_2}) & \Gamma_R(\mathbf{r_1}, \mathbf{r_2}) \\ \Gamma_R(\mathbf{r_1}, \mathbf{r_2}) & \Gamma_R(\mathbf{r_1}, \mathbf{r_2}) \end{bmatrix} \tag{9}$$

After inserting Eq. (9), the Eq. (8) is modified as,

$$\gamma^2(\mathbf{r_1}, \mathbf{r_2}) = \frac{\mathrm{tr}[\Gamma(\mathbf{r_1}, \mathbf{r_2})\, \Gamma^\dagger(\mathbf{r_1}, \mathbf{r_2})]}{|\mathrm{tr}[\Gamma(0)]|^2} = \frac{|\Gamma^O_{xx}(\mathbf{r_1}, \mathbf{r_2}) + \Gamma_R(\mathbf{r_1}, \mathbf{r_2})|^2 + 3\,|\Gamma_R(\mathbf{r_1}, \mathbf{r_2})|^2}{|\Gamma^O_{xx}(0) + 2\,\Gamma_R(0)|^2} \tag{10}$$

The first term in numerator of Eq. (10) is due to the interference of the object and reference random field which enables to retrieve the phase as well as to construct the complex coherence function of the object random field. This complex coherence function is used to retrieve the information hidden in the object random field whereas the second term in the numerator acts as a dc component.

3. **Experimental Details**

The schematic diagram of the experimental set up for the present study is shown in figure 1. A horizontally polarized He-Ne laser beam of wavelength 632.8 nm is cleaned and collimated using a combination of spatial filter, SF and lens, $L_1$ of 100 mm focal length. The collimated beam then enters a Mach-Zehnder interferometer (MZI) formed by beam splitters $BS_1$, $BS_2$ and mirrors $M_1$, $M_2$. The beam transmitted through beam splitter, $BS_1$ is folded by mirror, $M_1$ and is made to pass through an object "V" of size 7 mm x 6 mm and a ground glass, $GG_1$, which are kept at mutual separation of 200 mm. The combination of the object and $GG_1$ constitute the object arm of MZI. The speckle generated from $GG_1$ is referred as object speckle. On the other hand, the beam reflected from $BS_1$ is folded by mirror, $M_2$ and focused to an off-axis point on another ground glass, $GG_2$ using a microscopic objective, MO of magnification 10X and numerical aperture of 0.25. The beam, before entering the MO, is made diagonal using a half-wave plate, HWP. The combination of the HWP, MO and $GG_2$ constitute the reference arm of the MZI and the generated speckle is referred as reference speckle. The speckle generated from both the GGs are Fourier transformed by a lens, $L_2$ of 200 mm focal length and their far-field superposition is recorded by a charge coupled device, CCD camera, placed at the back focal plane of the lens, $L_2$. The average number of reference speckle in the superposed speckle pattern is controlled by changing the distance between the MO and $GG_2$. Sufficient amount of carrier frequency in the superposed speckle pattern is provided by adjusting the tilt angle of the MO.

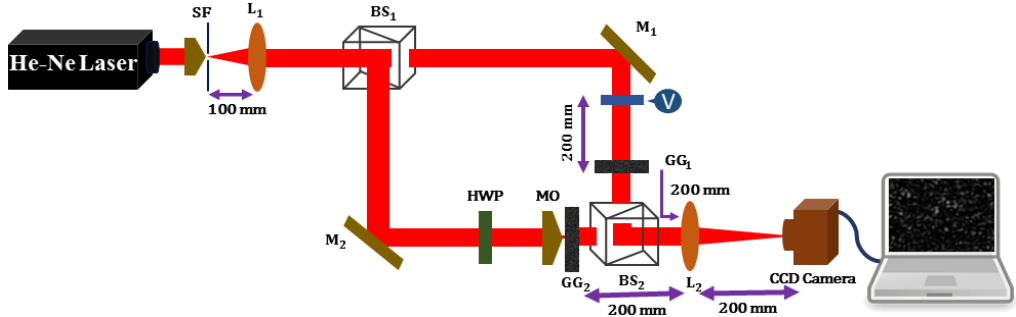

**Figure 1**. The schematic of the experimental set up

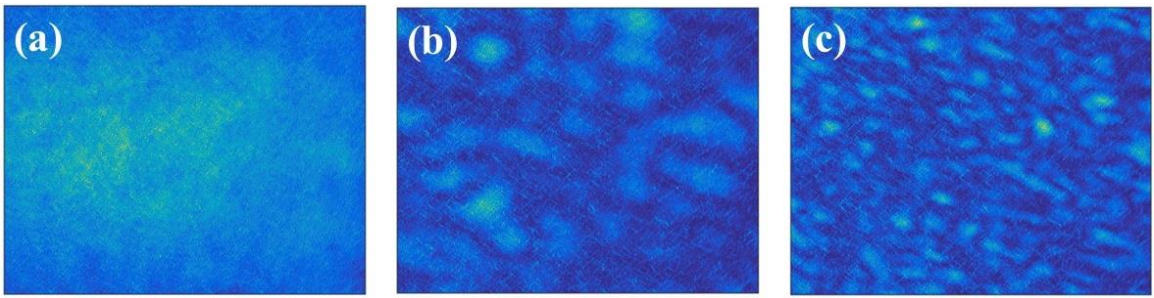

**Figure 2**. The recorded superposed far-field speckle patterns for average number of reference speckles: (a) 1, (b) 37, (c) 132

The effect of average number of reference speckles on the reconstructed object is studied by introducing reference speckles of different sizes in the superposed speckle pattern which is achieved by controlling the MO position and without giving any extra tilt angle to the MO. The superposed far-field speckle interferogram are recorded for different average number of reference speckles, of which three cases are shown in figure 2. The far-field reference speckles are also recorded by blocking the object beam for their characterization. The techniques used for the characterization of reference speckle and for the object information retrieval from the far-field speckle interferogram are discussed in detail in the next section.

## 4. Analysis and Result
The recorded far-field reference speckle patterns with different average number of speckles are characterized using the two-point intensity correlation function, discussed in Eq. (3). The average speckle size as well as the average separation between speckles are required to determine the average number of speckles present in the recorded speckle patterns. The full-width at half maxima (FWHM) of the far-field intensity correlation function is considered as the average speckle size. Whereas the average separation between the speckles is determined as the FWHM of intensity correlation function of the compliment of the speckle pattern which is obtained by taking compliment of the normalized speckle pattern i.e. by subtracting the normalized speckle pattern from unity. The effective area (including the separation between two speckles) covered by a speckle is determined from the estimated speckle size and the separation between the speckles. The average number of speckles present in the recorded speckle pattern is calculated by taking the ration of the area of recorded speckle pattern and the effective area covered by a speckle. When the number of speckles are very less (4 or less) in the recorded speckle pattern, it is very difficult to apply the discussed technique to find the speckle size as well as the average number of speckles. In such cases, the average number of speckles present in the recorded speckle pattern is counted manually. The maximum degree of correlation of the reference speckle, calculated using Eq. (3), is found to be changing with the average number of speckles and the variation is presented in figure 3.

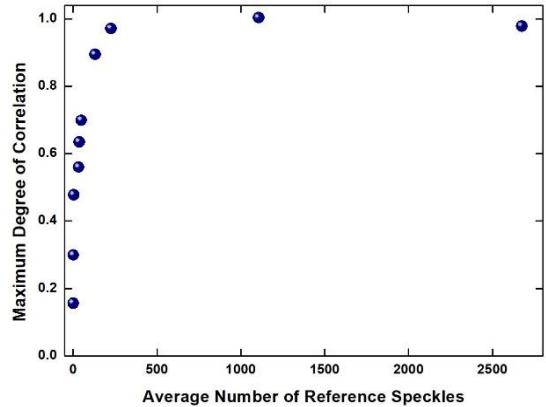

**Figure 3**. The change of the maximum degree of correlation with the average number of speckles for reference speckle

It is observed from figure 3 that the maximum degree of correlation of reference speckle changes from very close to zero to unity with the change of the average number of speckles. It is also found that the maximum degree of correlation is less (less than 0.71) for less average number of speckles (49 or less) and reaches to unity when the average number is high (226 or more). The observed changes in the maximum degree of correlation, even in case of a fully spatially polarized speckle, can be explained from the fact that when the average number of speckles are less, the statistics of the speckle deviates from the Gaussian failing to deliver the expected result i.e. maximum degree of correlation is unity. On the other hand, when the average number of reference speckle is high, the intensity distribution becomes fully random which follows the Gaussian statistics and the maximum degree of correlation is also found to be very near to the unity as expected in case of a fully spatially polarized speckle.

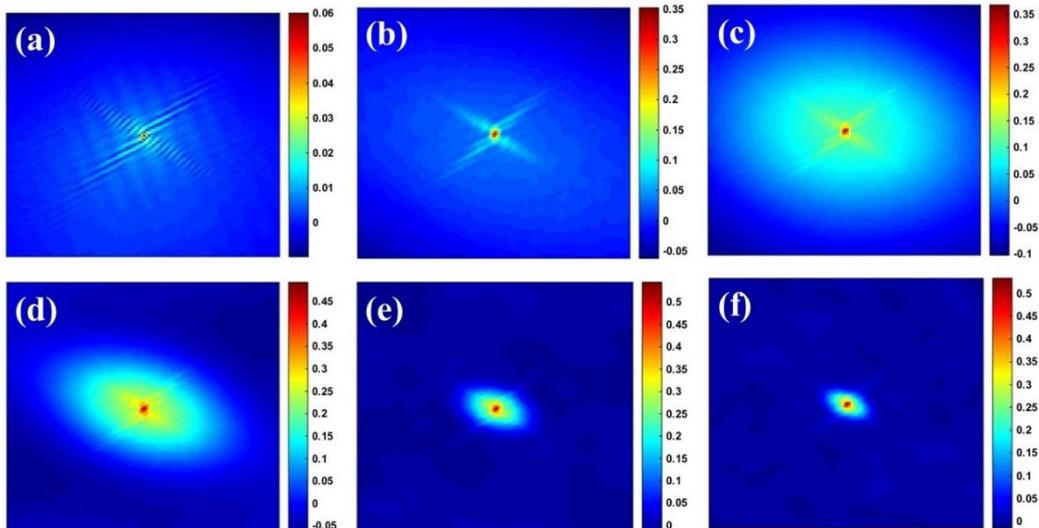

**Figure 4**. The coherence holograms for average number of reference speckles: (a) 1, (b) 4, (c) 37, (d) 132, (e) 1107, and (f) 2676

The intensity correlation technique, as discussed above, is also applied on the recorded resultant far-field speckle interferogram to construct the coherence hologram which is used for the object information retrieval. The coherence holograms obtained from intensity correlation for different average number of reference speckles are presented in figure 4. The presence of fringes in addition to a circular background is observed in the coherence holograms. The fringes in the coherence hologram is observed due to the correlation of the fringes present in the recorded interferogram, whereas the circular background is observed because of the correlation of the unmodulated speckles. It is observed from figure 4 that when the average number of reference speckles is less, the effect of correlation of the unmodulated speckles becomes weaker and that of the fringes becomes dominant in the coherence hologram (figures 4(a)-(b)). This is because in these cases almost all the object speckles interfere with the reference speckle and due to the less number of reference speckles, the correlation of non-interfering component (y-component) of the reference speckles is very low which can also be observed from figure 3. This results in dominant correlation fringes in the hologram. On the other hand, when the average number of reference speckle increases, the contribution of the correlation due to the fringes becomes weaker and that of the unmodulated speckles becomes stronger (figures 4(c)-(f)). This is observed because when the average number of reference speckles increase, the number of non-interfering reference as well as object speckles increase and the correlation of these unmodulated speckles reflects in the background of the fringe correlation in the coherence hologram. It is observed that the strength of correlation of the unmodulated speckles increases in proportion to the average number of reference speckles. The maximum of the colour bars presented along with the intensity correlation functions in figure 4, denote the squares of maximum degree of correlation which is found to be changing from very close to zero to more than 0.5.

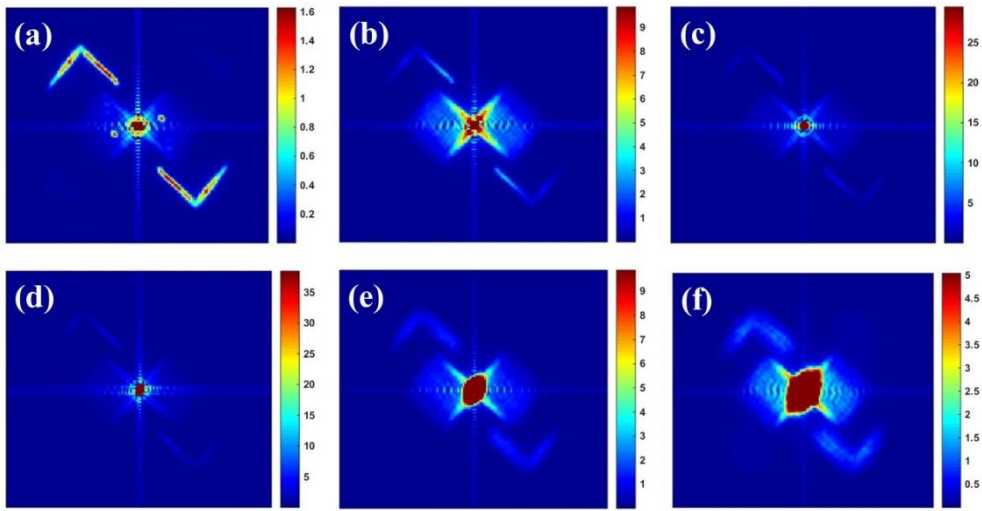

**Figure 5**. The retrieved objects for average number of reference speckles: (a) 1, (b) 4, (c) 37, (d) 132, (e) 1107, and (f) 2676

The object information scrambled into the object speckle is retrieved by applying the Fourier fringe analysis technique [28] on the fringes of the coherence hologram, obtained from the intensity correlation of the far-field superposed speckle. The retrieved objects for different average number of reference speckles are presented in figure 5. The maximum of the colour bars, presented along with each images in figure 5, denote the maximum values of the dc component present at the centre, which is treated as the maximum noise. It is observed from figure 5, that when the average number of reference speckles are lowest (figure 5(a)), the retrieved object quality is best and the quality starts to degrade with the increase in the average number of reference speckles. It is also found that when the average number of reference speckles are very high, the shape of the retrieved object is deformed (figures 5 (e-f)).

In order to study the effect of average number of reference speckles on retrieved object quality, each images in figure 5 is normalized with respect to the maximum noise. A rectangular window, enough to cover one of the retrieved objects present in the figure, is drawn around one object (here we have taken the object in the fourth quadrant) and sum of the normalized values inside the window, which is referred as the signal to noise ratio (SNR), is calculated for different average number of reference speckles. The variation of the SNR of the retrieved object with the average number of reference speckles is shown in figure 6. It can be observed that initially with the increase of average number of reference speckles, the SNR starts decreasing rapidly and reaches to a minimum value (when the average number of reference speckles is 49). And when the average number reference speckles is equal or more than 132, the SNR again starts to increase slowly with the increase of average number of reference speckles. It is also observed from figure 5 that in the latter case the shape of the recovered object starts to deform with the increase of average number of reference speckles.

A quick look at figure 3 reveals that when the average number reference speckles is equal or more than 132, the maximum degree of correlation reaches close to unity, indicating that the intensity distribution of reference speckle is Gaussian in nature in that region. Hence, the object retrieval process can be divided in two domains: Gaussian and non-Gaussian domain of the reference speckle. In the non-Gaussian domain, the SNR decreases sharply with the increase of the average number of reference speckles although the shape of the retrieved objects remains unchanged. Whereas in the Gaussian domain, the SNR increases with the increase of the average number of reference speckles and the shape of the retrieved object starts to deform.

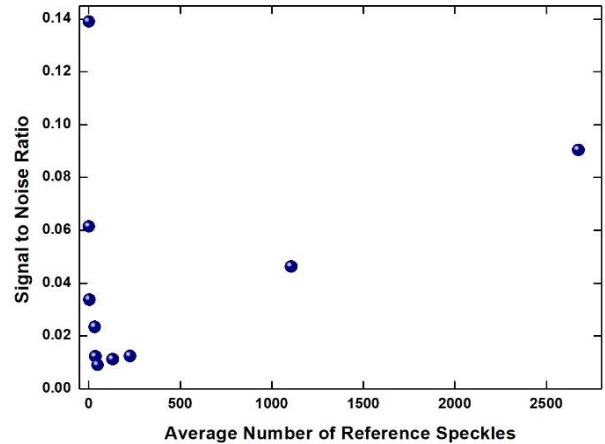

**Figure 6**. The change of SNR of the retrieved objects with average number of reference speckles

## 5. Conclusion
In this paper, we have studied the effect of average number of reference speckles on the SNR of the retrieved object in the off-axis holographic technique. It is shown that the average number of reference speckles play a major role in improving the retrieved object quality and the retrieval (both in terms of high SNR and shape) is best when the average number of reference speckles is unity and starts degrading with the increase of the number. This study will surely be helpful in providing a better insight in the off-axis holographic technique.

**Acknowledgments**
We would like to thank the Department of Science and Technology, Government of India and Indian Institute of Technology Kharagpur for their financial support towards carrying out the project.

**References**
[1] J. W. Goodman, Statistical Optics (Wiley, 2015).
[2] I. Freund, M. Rosenbluh, and S. Feng, Phys. Rev. Lett. **61**, 2328 (1988).
[3] J. V. Thompson, G. A. Throckmorton, B. B. Hokr, and V. V. Yakovlev, Opt. Lett. **41**, 1769 (2016).
[4] E. N. Leith, and J. Upatnieks, J. Opt. Soc. Am. **56**, 523 (1966).
[5] O. Katz, E. Small, Y. Guan, and Y. Silberberg, Optica **1**, 170 (2014).
[6] J. Bertolotti, E. G. van Putten, C. Blum, A. Lagendijk, W. L. Vos, and A. P. Mosk, Nature **491**, 232 (2012).
[7] O. Katz, P. Heidmann, M. Fink, and S. Gigan, Nat. Photonics **8**, 784 (2014).
[8] A. Porat, E. R. Andresen, H. Rigneault, D. Oron, S. Gigan, and O. Katz, Opt. Express **24**, 16835 (2016).
[9] H. M. Escamilla, E. R. Mendez, and D. F. Hotz, Appl. Opt. **32**, 2734 (1993).

[10] J. A. Newman, Q. Luo, and K. J. Webb, Phys. Rev. Lett. **116**, 073902 (2016).
[11] T. Wu, O. Katz, X. Shao, and S. Gigan, Opt. Lett. **41**, 5003 (2016).
[12] A. S. Marathay, Y. Hu, and L. Shao, Opt. Eng. **33**, 3265 (1994).
[13] S. M. Ebstein, J. Opt. Soc. Am. A **8**, 1442 (1991).
[14] D. A. Zimnyakov, V. V. Tuchin, R. A. Zdrajevsky, and Y. P. Sinichkin, Proc. of SPIE **3927**, 179 (2000).
[15] L. Tchvialeva, G. Dhadwal, H. Lui, S. Kalia, H. Zeng, D. I. McLean, and T. K. Lee, J. Biomed. Opt. **18**, 061211 (2013).
[16] D. Shi, S. Hu, and Y. Wang, Opt. Lett. **39**, 1231 (2014).
[17] Y. Zhu, J. Shi, Y. Yang, and G. Zeng, Appl. Opt. **54**, 1279 (2015).
[18] R. K. Singh, R. V. Vinu, and S. M. Anandraj, Opt. Eng. **53**, 104102 (2014).
[19] R. K. Singh, R. V. Vinu, and S. M. Anandraj, Appl. Phys. Lett. **104**, 111108 (2014).
[20] A. Roy, R. K. Singh, and M. M. Brundavanam, Appl. Phys. Lett. **109**, 201108 (2016).
[21] A. S. Somkuwar, B. Das, R. V. Vinu, Y. Park, and R. K. Singh, e-print arXiv:1511.04658 (2015).
[22] R. V. Vinu, and R. K. Singh, Opt. Lett. **40**, 1227 (2015).
[23] R. V. Vinu, and R. K. Singh, Appl. Phys. Lett. **109**, 111108 (2016).
[24] A. R. Ganesan, Proceedings of the National Seminar & Exhibition on Non-Destructive Evaluation 126 (2009).
[25] A. Roy, and M. M. Brundavanam, in 13th International Conference on Fiber Optics and Photonics, OSA Technical Digest, 2016, paper Tu4A.5.
[26] R. H. Brown, and R. Q. Twiss, Nature **177**, 27 (1956).
[27] J. Tervo, T. Setälä, and A. T. Friberg, Opt. Express **11**, 1137 (2003).
[28] M. Takeda, H. Ina, and S. Kobayashi, J. Opt. Soc. Am. 72,156 (1982).